# Advancements in Feature Extraction Recognition of Medical Imaging Systems Through Deep Learning Technique

Qishi Zhan[1], Dan Sun[2], Erdi Gao[3], Yuhan Ma[4], Yaxin Liang[5], Haowei Yang[6]

[1]Marquette University,USA,qishizhan7@gmail.com

[2]Washington University in St. Louis,USA,sun.dan@wustl.edu

[3]New York University,USA,ge2093@nyu.edu

[4]Johns Hopkins University,USA,yma62@jh.edu

[5]University of Southern California,USA,yaseen.liang@outlook.com

[6]University of Houston,USA,yanghaowei09@gmail.com

*Abstract*—This study introduces a novel unsupervised medical image feature extraction method that employs spatial stratification techniques. An objective function based on weight is proposed to achieve the purpose of fast image recognition. The algorithm divides the pixels of the image into multiple subdomains and uses a quadtree to access the image. A technique for threshold optimization utilizing a simplex algorithm is presented. Aiming at the nonlinear characteristics of hyperspectral images, a generalized discriminant analysis algorithm based on kernel function is proposed. In this project, a hyperspectral remote sensing image is taken as the object, and we investigate its mathematical modeling, solution methods, and feature extraction techniques. It is found that different types of objects are independent of each other and compact in image processing. Compared with the traditional linear discrimination method, the result of image segmentation is better. This method can not only overcome the disadvantage of the traditional method which is easy to be affected by light, but also extract the features of the object quickly and accurately. It has important reference significance for clinical diagnosis.

## I. Introduction

Under the trend of digital, intelligent, and humanized development of medical diagnosis technology, imaging diagnosis methods based on magnetic resonance, CT, ultrasound and other medical instruments have been widely adopted[1-4]. Image analysis technology is employed to extract patient-relevant information, whereas image processing technology facilitates the rapid and effective segmentation of features. These technologies provide a crucial foundation for doctors to formulate consultation, surgical, and diagnostic plans[5]. In addition, due to the complexity, irregularity and individual differences of each organ structure of the human body, effectively extracting the characteristics of objects from the image is the key to improving the diagnosis and treatment of diseases. The Gaussian curve method, direction template method, matrix method, and threshold method are the most commonly used methods at present. Some researchers have proposed a new idea of Gaussian fitting subpixel boundary extraction, that is, first select a set of points at the boundary of the original image, carry out gray-level steps on them, and then carry out high-precision fitting [6-7]. This method exhibits strong resistance to sensitivity and is suitable primarily for cases involving slowly varying normal vectors. Previous studies have demonstrated that the long-stroke gait recognition method, which relies on multi-shape template reinforcement[8-10], can construct the target orientation template. This is achieved by extracting the transverse, longitudinal, shape distribution, and movement distribution templates of the target[11]. With its good anti-noise performance, the target can be better divided into more detailed targets. However, in practical applications, its accuracy rate is difficult to meet the needs of medical diagnosis [12]. Some studies have proposed a boundary detection method based on the Hessian matrix, that is, the Hessian matrix is used to locate the boundary, eliminate the rough and redundant background texture of the boundary, and achieve boundary smoothing and continuity smoothing [13]. Compared with the above methods, the threshold method is fast, diverse and more flexible, which is of great development and research significance.

The project focuses on current popular image processing methods as its research object and explores the integration of contemporary medical imaging features, emphasizing real-time efficiency and precision. This is achieved through a spatial hierarchical approach. In each subdomain, the threshold of each subdomain is determined by constructing an objective function with weights [14]. The algorithm first divides the image into multiple subdomains and then uses the quadtree method to access each subdomain at multiple levels. Finally, the simplex method is introduced to optimize the sub-image thresholds without constraints, This approach significantly enhances the precision and expedience of image recognition processes, offering substantial reference significance for clinical medical diagnoses.

## II. SYSTEM ALGORITHM

This algorithm does not need to differentiate the objective function and has the characteristics of simple structure and intuition. In addition, because the threshold is a positive integer in the range (0,255), the integer requirement for the threshold can be ignored when solving this problem [15]. For the best scheme that cannot reach the integer type, the method of "rounding" is used to solve, and the result is close to the best scheme. The main implementation process of the algorithm is shown in Figure 1 (the picture is quoted in A framework for feature extraction in risk prediction).

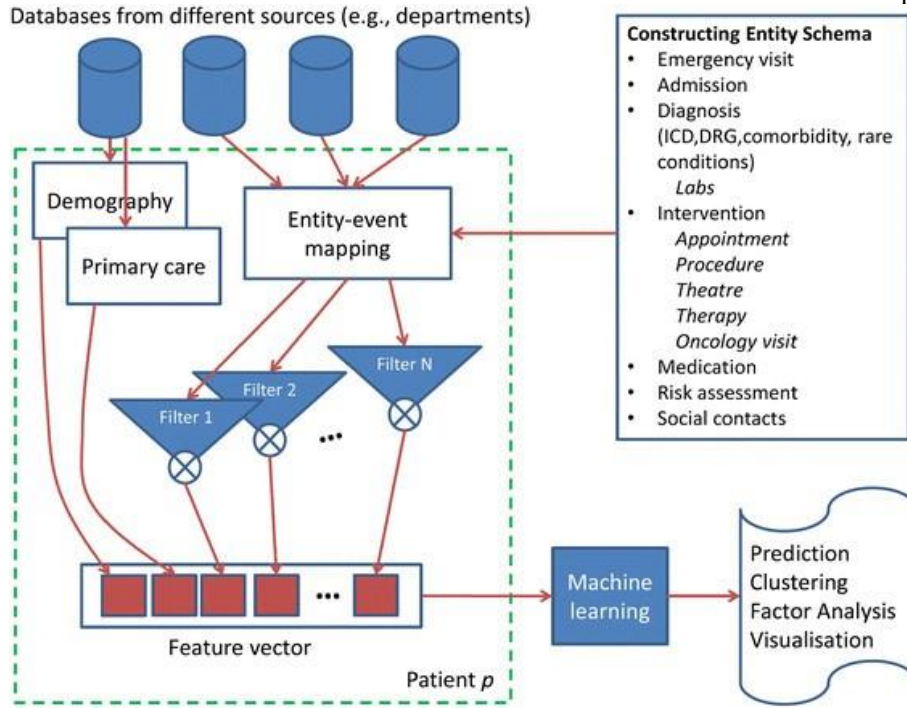

Fig. 1. Algorithm logic diagram

The algorithm adopts the weight method to dynamically adjust the objective function based on synthesizing the complexity and other characteristics of the medical image and adopts the iterative method to obtain the best result [16]. This detection technology not only has a good ability to adapt to the surrounding situation[17-18] but also can make the detection results real-time and accurate, which can provide a reference for clinicians[19].

## III. GENERALIZED DISCRIMINANT ANALYSIS FEATURE EXTRACTION

Initially, the non-linear function is mapped into a higher-dimensional feature space, followed by linear discrimination. To circumvent the direct explicit manipulation of the transformed data, a kernel-based approach is adopted to determine the sample values within the feature space[20]. Let $\lambda_1, \lambda_2, \cdots, \lambda_m$ be the class of $Z$ samples and the original sample $u$ be the $n$ dimensional real vector $u \in R^n$. After nonlinear mapping $\phi$, the corresponding sample vector is $\phi(u) \in G$. The in-class discrete matrix $(D_w^\phi)$, the inter-class discrete matrix $(D_b^\phi)$ and the overall discrete matrix $(D_t^\phi)$ of the training sample in the high-dimensional feature space $G$ are respectively

$$D_w^\phi = \frac{1}{M} \sum_{i=1}^{Z} \sum_{j=1}^{M_i} (\phi(u_j^i) - n_i^\phi)(\phi(u_j^i) - n_i^\phi)^T \quad (1)$$

$$D_b^\phi = \frac{1}{M} \sum_{i=1}^{Z} \frac{M_i}{M} (n_i^\phi - n_0^\phi)(n_i^\phi - n_0^\phi)^T \quad (2)$$

$$D_t^\phi = \frac{1}{M} \sum_{j=1}^{M} (\phi(u_j) - n_0^\phi)(\phi(u_j) - n_0^\phi)^T \quad (3)$$

$n_i^\phi = E\{\phi(u) | \lambda_i\}$ is the mean of class $i$ training samples in feature space $G$.

$$H_1(\kappa) = \frac{\kappa^T D_b^\phi \kappa}{\kappa^T D_w^\phi \kappa} \quad (4)$$

Where $\kappa$ is any non-zero column vector. In the feature space $G$, GDA seeks to find a set of discriminant vectors $\kappa_1, \cdots, \kappa_d$ that satisfy the Fisher criterion function (4) while maximizing it.

The first discriminant vector of GDA is taken as Fisher's best discriminant vector. The eigenvector $\kappa_1$ corresponding to the maximum eigenvalue of the generalized eigenequation $D_b^\phi \kappa = \eta D_w^\phi \kappa$ is used to obtain the first $s$ discriminant vectors $\kappa_1, \cdots, \kappa_s$ of GDA. Then the $s+1$ discriminant vector $\kappa_{s+1}$ can be obtained by solving the following optimization problems:

$$I \begin{cases} \max(H_1(\kappa)) \\ \kappa_j^T \kappa = 0, j = 1, \cdots, s \\ \kappa \in G \end{cases} \quad (5)$$

$$\kappa = \sum_{i=1}^{M} \sigma^i \phi(u_i) = \phi \sigma \quad (6)$$

$\phi = (\phi(u_1), \cdots, \phi(u_M)); \sigma = (\sigma^1, \cdots, \sigma^M)^T, \sigma$ is the best discriminant direction $\kappa$ in feature space $G$. The sample $\phi(u)$ in the feature space $G$ is projected onto $\kappa$

$$\kappa^T \phi(u) = \kappa^T \phi^T \phi(u) = \sigma^T \mu_u \quad (7)$$

Corresponding to original sample $u \in R^n, \mu_u$ are $M$ nuclear samples corresponding to original training sample $u_1, u_2, \cdots, x_M$. Vector then the kernel matrix is $U = (\mu_{u_1}, \mu_{u_2}, \cdots, \mu_{u_M})$. The mean vector of the training sample class and the population mean vector in the feature space $G$ are projected onto $\kappa$ respectively, then

$$\kappa^T n_i^\phi = \sigma^T \phi^T \frac{1}{M_i} \sum_{i=1}^{M_i} \phi(u_k^i) = \sigma^T \delta_i \quad (8)$$

$$\kappa^T n_0^\phi = \sigma^T \phi^T \frac{1}{M} \sum_{i=1}^{M} \phi(u_k^i) = \sigma^T \delta_0 \quad (9)$$

Among them:

$$\delta_i = \begin{pmatrix} \frac{1}{M_i} \sum_{k=1}^{M_i} (\phi(u_1) \cdot \phi(u_k^i)), L \\ \frac{1}{M_i} \sum_{k=1}^{M_i} (\phi(u_M) \cdot \phi(u_k^i)) \end{pmatrix} \quad (10)$$

$$\delta_0 = \begin{pmatrix} \frac{1}{M} \sum_{k=1}^{M} (\phi(u_1) \cdot \phi(u_k^i)), L \\ \frac{1}{M} \sum_{k=1}^{M_i} (\phi(u_M) \cdot \phi(u_k^i)) \end{pmatrix} \quad (11)$$

According to formula (8), (10) and (11)

$$\kappa^T D_b^f \kappa = \sigma^T U_b \sigma \quad (12)$$

$$\kappa^T D_w^f \kappa = \sigma^T U_w \sigma \quad (13)$$

$$\kappa^T D_t^f \kappa = \sigma^T U_t \sigma \quad (14)$$

Among them:

$$U_b = \sum_{i=1}^{Z} \frac{M_i}{M} (\delta_i - \delta_0)(\delta_i - \delta_0)^T \quad (15)$$

$$U_w = \frac{1}{M} \sum_{i=1}^{Z} \sum_{j=1}^{M_i} (\mu_{u_j^i} - \delta_i)(\mu_{u_j^i} - \delta_i)^T \quad (16)$$

$$U_t = \frac{1}{M} \sum_{j=1}^{M} (\mu_{u_j} - \delta_i)(\mu_{u_j} - \delta_i)^T \quad (17)$$

$U_b, U_w, U_t$ in equations (15) to (17) is called inter-class discrete matrix, intra-class discrete matrix and, respectively. The global discrete matrices are all $M \times M$ nonnegative definite symmetric matrices.

$$H_1'(\sigma) = \frac{\sigma^T U_b \sigma}{\sigma^T U_w \sigma} \quad (18)$$

Where $\sigma$ is any $M$ dimensional nonzero column vector. The orthogonal constraints in high dimensional eigenspace $G$ are equivalent to

$$\kappa_i^T \kappa_i = \sigma_i^T \phi^T \phi_j \sigma_j = \sigma_i^T U \sigma_j = 0,$$
$$\forall i \neq j; i, j = 1, \cdots, d$$

Then model $I$ represented by a kernel matrix is equivalent to

$$\begin{cases} \max(H_1'(\sigma) \\ \sigma_j^T U \sigma = 0, j = 1, \cdots, s \\ \sigma \in R^M \end{cases} \quad (19)$$

If the first $s$ maximization criterion function (18) satisfies the orthogonal constraint condition. After the best kernel discriminant vector $\sigma_1, \cdots, \sigma_s$ is obtained, the $s+1$ best kernel discriminant vector $\sigma_{s+1}$ can be determined by model II. The first best kernel discriminant direction $\sigma_1$ is the generalized eigenequation $U_b \sigma = \eta U_w \sigma$. Model I and Model II of GDA are formally equivalent. If $\{\sigma_1, \sigma_2, \cdots, \sigma_d\}$ is the best kernel discriminant vector set determined by model II, then the best discriminant vector set for model I is $\{\kappa_1, \kappa_2, \cdots, \kappa_d\}$.

IV. ALGORITHM SOFTWARE IMPLEMENTATION

The number of subfields is obtained through iteration and then input into the optimization problem. Then the simplex algorithm is used for optimization. The optimal threshold value is obtained by objective function and simplex method, which provides an effective method for extracting medical image features accurately.

Although the human body has similar structure, it is affected by many factors such as nature and environment, and there are significant differences in organizational characteristics between them. The brain CT scanning was performed on 50 patients in Jordan University Hospital (JUH) [21]. The dataset is handled through the Linked Data methodology to integrate various data formats, crucial in academic studies[22]. This organized approach allows for the interconnection of information, boosting the ability of different images to work together. This feature is particularly valuable in the realms of machine learning and artificial intelligence, where the integrity of data is essential for effectively training models and securing precise outcomes. Experimental conditions: multi-layer spiral scanning machine was used, the scanning layer was 5 mm high, and the scanning layer was 10 layers from the cranial top to the cranial base, including cerebral infarction, cerebral hemorrhage and brain tumor. Through the analysis of 100 CT scan data, the paper obtained 100 CT scan data from simple cranial roof to extremely complex cranial skull, and obtained satisfactory results. The disadvantages are: 1) Because the distance between the extracranial body and the skull is very close, resulting in the loss of the skull and cannot be recognized. 2) There are also some eye structures in the head CT images, but this situation has little impact on the identification results. Figure 2 shows stratified brain CT sampling.

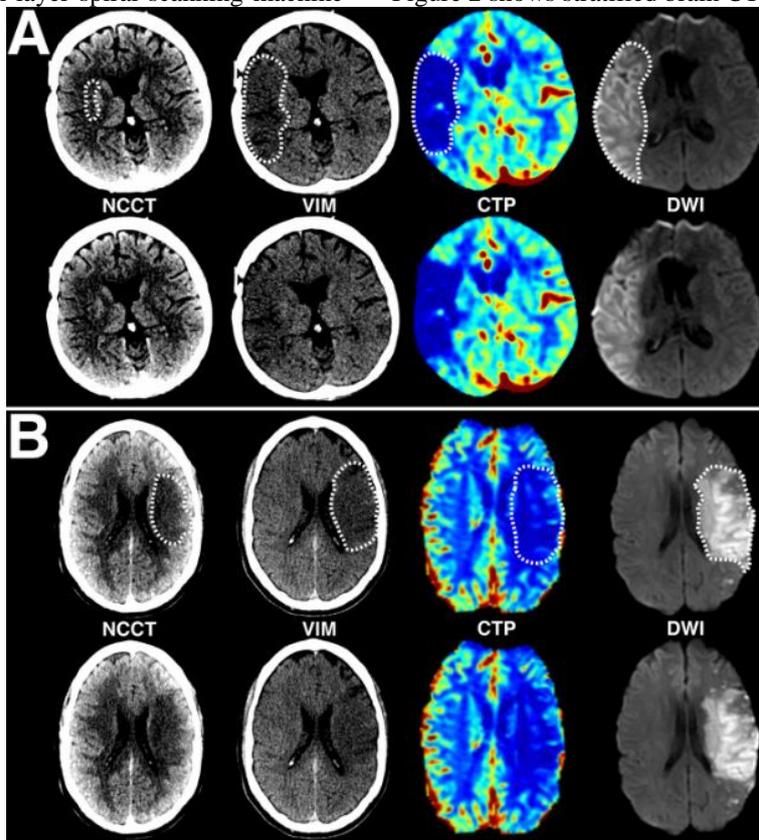

Fig. 2. Craniocerebral CT stratified sampling

The BET method is used to compare the method. A method of CT image classification based on unrestricted weights is proposed [23]. By adjusting the weights of each index, good results are obtained. However, when the BET method is used for image identification and corresponding parameters, the recognition results are shown in Figure 3. The indicators and results of the comparative trial are shown in Table 1. Under the same test environment, this method has better anti-noise and anti-noise performance than BET method . Meanwhile, hierarchical memory structure and non-restrictive features are used to process it, which can significantly improve image access efficiency and identification accuracy (Figure 4).

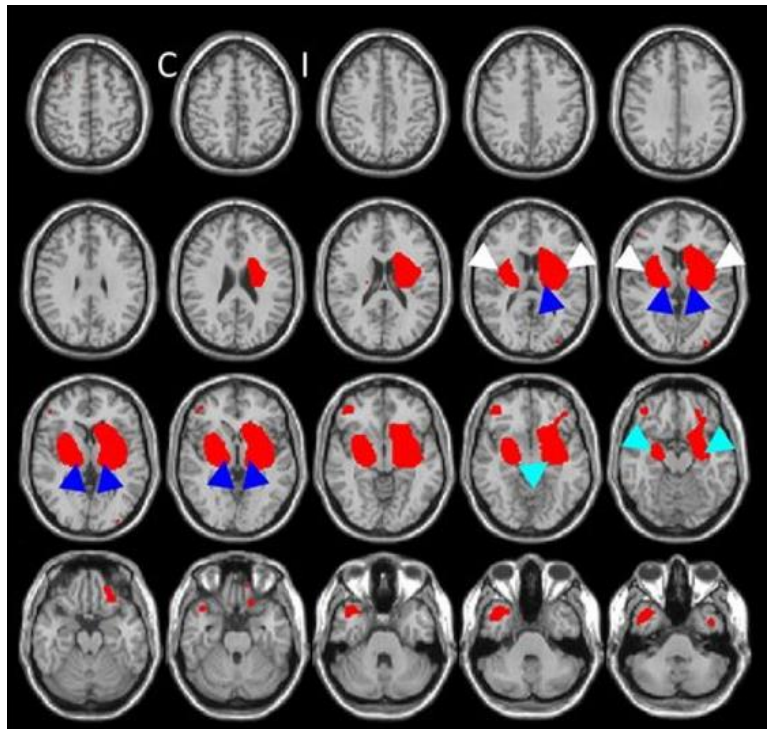

Fig. 3. Effect of skull base recognition

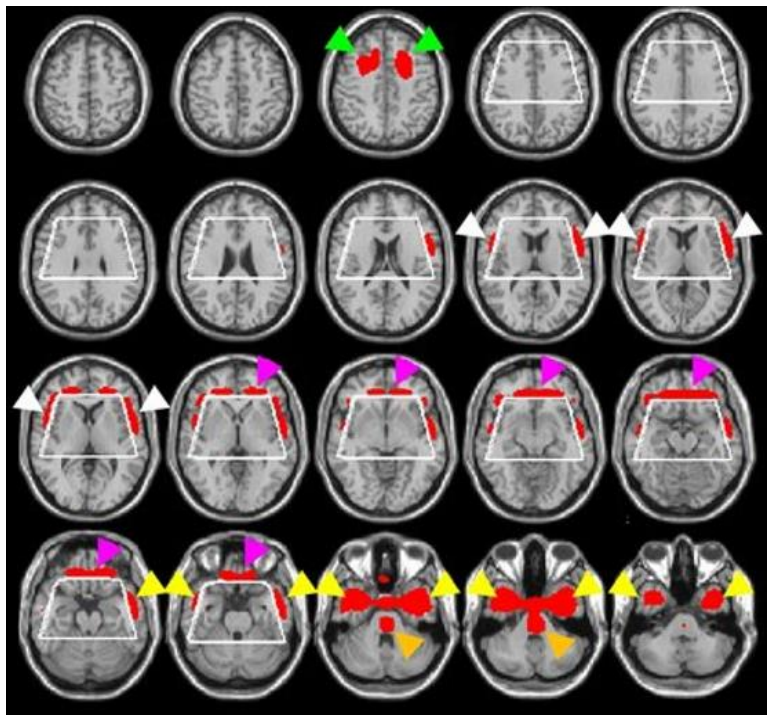

Fig. 4. Effect diagram of skull base recognition (BET algorithm)

TABLE I. COMPARISON OF ALGORITHMS

| Experimental method | Spatial stratification | Binding character | Distortion degree | Reliability |
| --- | --- | --- | --- | --- |
| This algorithm | quartering | unconstraint | <0.2 | 0.96 |
| BET algorithm | No layering | Be limited | >0.3 | 0.85 |

The experimental results show that the research station can subdivide the whole three-dimensional space according to the characteristics of medical images, and the weighted coefficient objective function can improve the segmentation effect from the perspective of image spatial structure and light sensitivity.

## V. CONCLUSION

According to the characteristics of medical images, an adaptive detection method based on unrestricted conditions is presented. This method not only makes full use of the role of environmental factors but also makes full use of the complex characteristics of human tissues. A recursive spatial image segmentation method based on four access methods is proposed, and the optimal threshold of the image is obtained by unrestricted adaptive optimization objective function. The separation test and feature analysis of multiple human organs were conducted to create a comprehensive image-characteristic database of human tissues. This database serves as a critical source of data for advancing future pattern recognition research.